\documentclass[onecollarge,natbib,final]{svjour2}

\bibpunct{[}{]}{;}{n}{}{,} 
\smartqed  
\usepackage{graphicx}
\usepackage{color}
\usepackage{amsmath,amssymb,bm}
\usepackage{amsfonts}
\usepackage{tensor}
\usepackage{dcolumn}
\usepackage{units}

\newcommand{\paper}{Contribution}
\newcommand{\paren}[1]{\ensuremath{\left(#1\right)}}
\newcommand{\ket}[1]{\left| #1 \right>} 
\newcommand{\braket}[2]{\left \langle #1 \vphantom{#2} \right|
 \left. #2 \vphantom{#1} \right\rangle} 
\newcommand{\matrixel}[3]{\left\langle #1 \vphantom{#2#3} \right|
 #2 \left| #3 \vphantom{#1#2} \right\rangle} 
\renewcommand{\sp}{s.p.}
\journalname{Few-Body Systems}
\begin{document}

\title{Strongly interacting few-fermion systems in a trap
}


\author{Christian Forss\'en  \and
  Rikard Lundmark \and
  Jimmy Rotureau \and
  Jonathan Larsson \and
  David Lidberg
}


\institute{C. Forss\'en \at
              Department of Fundamental Physics, Chalmers University
              of Technology, SE-412 96 G\"oteborg, Sweden \\ 
              Department of Physics and Astronomy, University of
              Tennessee, Knoxville, TN 37996, USA\\
              Physics Division, Oak Ridge National Laboratory, Oak Ridge,
              TN 37831, USA \\
              \email{christian.forssen@chalmers.se}           
           \and
           R. Lundmark \and  J. Rotureau \and  J. Larsson \and
           D. Lidberg \at
           Department of Fundamental Physics, Chalmers University
           of Technology, SE-412 96 G\"oteborg, Sweden 
}

\date{Received: date / Accepted: date}

\maketitle

\begin{abstract}
  Few- and many-fermion systems on the verge of stability, and
  consisting of strongly interacting particles, appear in many areas
  of physics. The theoretical modeling of such systems is a very
  difficult problem.
  In this work we present a theoretical framework that is based on the
  rigged Hilbert space formulation. The few-body problem is solved by exact
  diagonalization using a basis in which bound, resonant, and
  non-resonant scattering states are included on an equal footing.
  Current experiments with ultracold atoms offer a fascinating
  opportunity to study universal properties of few-body systems
  with a high degree of control over parameters such as the
  external trap geometry, the number of particles, and even the
  interaction strength. In particular, particles can be allowed to
  tunnel out of the trap by applying a magnetic-field gradient that
  effectively lowers the potential barrier. The result is a tunable
  open quantum system that allows detailed studies of the tunneling
  mechanism.
  In this \paper{} we introduce our method and present results for the
  decay rate of two distinguishable fermions in a one-dimensional trap
  as a function of the interaction strength. We also study the
  numerical convergence.
  Many of these results have been previously published (R. Lundmark,
  C. Forss\'en, and J. Rotureau, arXiv: 1412.7175). However, in this
  \paper{} we present several technical and numerical details of our
  approach for the first time.
\end{abstract}

\section{Introduction}
\label{intro}
%
The tunneling of particles, energetically confined by a potential
barrier, is a fascinating quantum phenomenon which plays an important
role in many physical systems. An exciting recent development in the
context of multiparticle tunneling is the experimental realization
of few-body Fermi systems with ultracold
atoms~\cite{Serwane:2011hp,2013PhRvL.111q5302Z}. These setups are
extremely versatile as they are associated with a high degree of
experimental control over key parameters such as the number of
particles and the shape of the confining potential. In addition, the
interaction between particles can be tuned using Feshbach
resonances~\cite{Chin:2010kl}, which in the case of trapped particles
turns into a confinement-induced
resonance~\cite{Olshanii:1998jr,Zuern:2013cr}. The resulting
interparticle interaction is of very short range compared to the size
of the systems, and can be modeled with high accuracy by a zero-range
potential.
Such tunable open quantum systems provides a unique opportunity to
investigate the mechanism of tunneling as a function of the trap
geometry and the strength of the interparticle interaction.

\begin{figure}
\centering
\includegraphics[width=0.6\columnwidth]{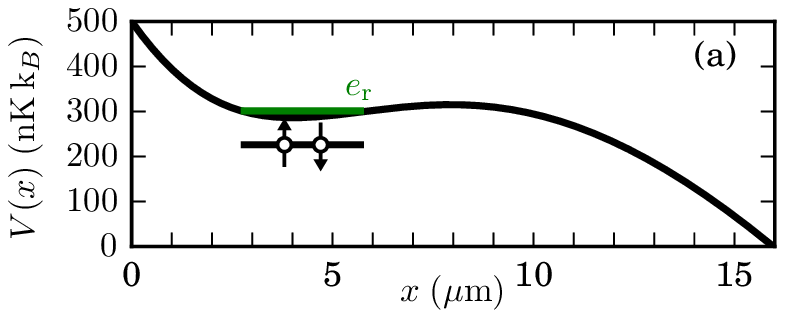}\\[-2ex]
\includegraphics[width=0.6\columnwidth]{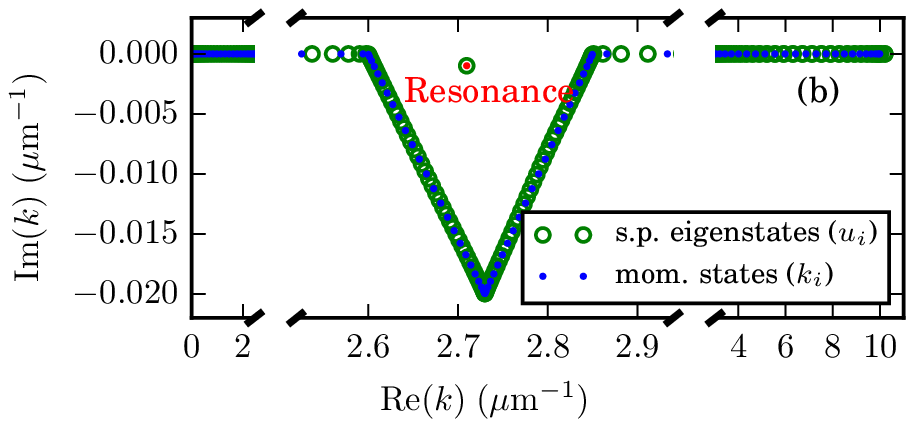}
\caption{\label{fig:OneDContour}Panel (a): Trap potential, indicating
  the position of \sp{} and two-body resonance states. Panel (b):
  Complex-momentum contour and Berggren basis states, highlighting the
  position of the \sp{} resonance pole.} 
\end{figure}

In this \paper{} we will consider a system of interacting,
two-component fermions in a finite-depth potential trap.  The trap is
not deep enough to support a single-particle (\sp{}) bound state, but
does provide a quasi-bound state with a finite lifetime. We employ an
effective 1D potential corresponding to the stated potential for the
experimental setup in Ref.~\cite{2013PhRvL.111q5302Z}
\begin{align}
V(x) = p V_0 \paren{1-\frac{1}{1 + \paren{\frac{x}{x_R}}^2 }}
-c_{B,\sigma}\mu_B B' x. \label{eq:Vexp},
\end{align}
where $p V_0$, $x_R$, and $B'$ are tunable parameters and $c_{B
  ,\sigma} \approx 1$. This potential is illustrated in
Fig.~\ref{fig:OneDContour}(a).

We recently introduced the rigged Hilbert space approach to the study
of such open quantum systems~\cite{2014arXiv1412.7175L}.  This method
extends beyond the domain of Hermitian quantum mechanics and includes
also time-asymmetric processes such as decays (see e.g.\
Ref.~\cite{delaMadrid:2005gt} and references therein).
In nuclear physics this formulation has been
employed in the Gamow Shell Model~\cite{Michel:2002hk,
  Rotureau:2006fo, Hagen:2006fy,Michel:2009jg,Papadimitriou:2011ik,
  2013FBS....54..725R} to study threshold states and decay
processes. 
Recently, it has also been used to model near-threshold, bound states
of dipolar molecules~\cite{Fossez:2013fj}.
\section{Method}
\label{sec:method}
We will obtain solutions of the many-body Hamiltonian for interacting
particles using an expansion of \sp{} states in the so-called Berggren
basis~\cite{Berggren:1968zz}. This complex-momentum basis includes
S-matrix poles (bound and resonant states) as well as non-resonant
scattering states. The use of this basis, constituting a rigged
Hilbert space, is key to our approach as it allows to consistently
include the continuum when finding eigensolutions of the open quantum
system. The corresponding completeness relation is a generalization of
the Newton completeness relation~\cite{1960JMP.....1..319N} (defined
only for real energy states) and reads
\begin{align}
\sum_n |u_n\rangle \langle \tilde{u}_n|+\int_{L^+} dk | u_k\rangle \langle
\tilde{u}_k|=1, \label{eq:berg_comp} 
\end{align}
where $|u_n\rangle$ correspond to poles of the S-matrix, and the
integral of states along the contour $L^+$, extending below the
resonance poles in the fourth quadrant of the complex-momentum plane,
represents the contribution from the non-resonant scattering
continuum~\cite{Berggren:1968zz}. This basis also features a
non-conjugated inner product.
%
\paragraph{Single-particle basis}
%
In order to generate the \sp{} Berggren basis to be used in the
many-body calculation, we start with a complex-momentum basis
\begin{equation}
\braket{x}{k,\Pi} = \left\{ \begin{array}{ll}
\sqrt{\frac{2}{\pi}} \sin(kx) & (\Pi=0) \\
\sqrt{\frac{2}{\pi}} \cos(kx) & (\Pi=1)
\end{array} \right.
\end{equation}
that is equivalent to a combination of left- and right-travelling
plane waves. In fact, this set corresponds to the Berggren basis for
a \sp{} Hamiltonian with $V=0$. In practice, the momentum integral of
Eq.~\eqref{eq:berg_comp} is performed using Gauss-Legendre quadrature and
the momentum basis states are discretized accordingly.

We will use the short-hand notation $\ket{k_i}$ to denote an eigenstate with
complex momentum $k_i$ and parity $\Pi_i$. The matrix representation
\begin{equation}
h_{ij}  = \matrixel{k_i}{h}{k_j}
\end{equation}
of the one-body Hamiltonian%
\begin{equation}
h(x) = -\hbar^2 / (2m) d^2 / dx^2 + V(x)
\label{eq:H1}
\end{equation}
is in general not Hermitian and not symmetric. The latter property can
be recovered by redefining
$\tilde{h}_{ij} = \sqrt{\frac{w_i}{w_j}}h_{ij}$, 
where $w_{i,j}$ are the Gauss-Legendre weights. The complex-momentum
contour $L^+$ is selected such that the \sp{} resonance energy, which is one
of the eigenvalues, appears above the contour in the fourth quadrant. A
specific example is shown in Fig.~\ref{fig:OneDContour}(b), in which
the contour $L^{+}$ consists of four segments and is truncated at $k =
  k_\mathrm{max}$.
  The \sp{} resonance eigenstate $|u_\mathrm{res}\rangle$ can be
  described as a Gamow state~\cite{1928ZPhy...51..204G}. Such a state
  behaves asymptotically as an outgoing wave with a complex-energy
  $e_\mathrm{res}=e_{r}-i \gamma_r / 2$. The imaginary part of the
  energy corresponds to the decay width $\gamma_r$ and gives the
  half-life of the \sp{} state, $t_{1/2}=\ln{(2)} \hbar / \gamma_r$,
  and the \sp{} tunneling rate $\gamma_1 = \gamma_r / \hbar$. The full
  set of eigenvectors to the \sp{} Hamiltonian~\eqref{eq:H1} includes
  the resonance state, as well as non-resonant scattering states with
  complex momenta very close to the original contour. Together, these
  eigenstates correspond to our \sp{} basis states $U_1 \equiv
  \{|u_i\rangle\}$~\cite{Michel:2009jg}.
%
\paragraph{Many-body problem}
%
The interaction between fermionic atoms in different hyperfine states
is modeled by the zero-range potential $V_{12}(x_1,x_2)=g
\delta\paren{x_1 - x_2}$, with $g$ the tunable interaction strength.  The
fermions will be referred to according to their hyperfine spin state as
``spin-up'' ($\uparrow$) and ``spin-down'' ($\downarrow$), thus making
an obvious connection with systems of spin 1/2 particles
(e.g. electrons or nucleons).

We will consider two particles in the trap, being the simplest
example of a many-body system. The Hamiltonian is
\begin{align}
H(x_1,x_2) = \sum_{i=1}^{2} \left [ -\frac{\hbar^2}{2m} \frac{d^2}{dx_i^2} +
  V(x_i)\right ]+g \delta(x_1-x_2). \label{eq:H2} 
\end{align}
The two-particle basis $T_2$ is then naturally constructed from the
\sp{} bases for the spin-up and -down fermions as $T_2 \equiv
U_1(\uparrow) \otimes U_1(\downarrow)$. Note that the spin-dependent
$c_{B,\sigma}$-term in the trap potential~\eqref{eq:Vexp} will, in
general, result in different \sp{} bases for different spin states. In
the following we will not directly compare to experimental results
and will therefore restrict ourselves to spin-independent trap
potentials with $c_{B,\sigma} = 1$.
Let us first consider the situation of two non-interacting particles,
i.e. $g=0$.  In this case, the ground state of the system,
$|\Phi^{(0)}\rangle$, corresponds to the two distinguishable fermions
both occupying the resonant (quasi-bound) state $|u_\mathrm{res}
\rangle$. In this configuration, both particles are localized in the
trap for a finite amount of time, before tunneling out through the
potential barrier. 

In order to construct the many-body Hamiltonian matrix we first need
to evaluate the interaction matrix elements between Berggren states.
Note that the basis functions along the complex contour diverge for $x
\rightarrow \infty$. As a consequence, the matrix elements of the
two-body interaction are not finite in the Berggren basis. We solve
this issue by regularizing the two-body matrix elements between states
in $T_2$ using an expansion in the harmonic oscillator (HO)
basis~\cite{Hagen:2006fy}
\begin{equation}
\matrixel{u_{i},u_{j}}{V_{12}}{u_{l},u_{m}} = 
\sum_{\begin{array}{l} n_\alpha, n_\beta \\ n_\gamma, n_\delta \end{array}}^{n_\mathrm{max}} \braket{u_{i}}{n_{\alpha}}\braket{u_{j}}{n_{\beta}}
\braket{u_{l}}{n_{\gamma}}\braket{u_{m}}{n_{\delta}}
\matrixel{n_{\alpha},n_{\beta}}{V_{12}}{n_{\gamma},n_{\delta}},
\label{eq:tbme}
\end{equation}
where the sum runs over all HO states up to some truncation
$n_\mathrm{max}$. 
In the end, our Hamiltonian~\eqref{eq:H2} matrix in this rigged Hilbert
space will be non-Hermitian, but complex symmetric.  The spectrum will
include bound, resonant and scattering many-body states.

\paragraph{Diagonalization} 
%
We will now turn to the problem of finding the resonance state in the
eigenspectrum of the many-body Hamiltonian matrix.
While there are several algorithms available for finding extreme
eigenvalues of Hermitian matrices, our problem is different.
We are searching for many-body resonance solutions,
$|\Phi_\mathrm{res}\rangle$, that are characterized by
outgoing boundary conditions and a complex energy
$E_\mathrm{res}=E_r-i \Gamma_r / 2$, where $\Gamma_r$ is the decay
width due to the emission of particles out of the trap.
In general, these physical states will correspond to specific complex
eigenvalues in the interior part of the eigenvalue spectrum.
Such eigenstates can be identified by the property that they will be
independent of the particular choice of ${L^+}$ as long as the
Berggren completeness relation~\eqref{eq:berg_comp} holds, i.e.\
$k_\mathrm{max}$ and the number of discretization points both need to
be large enough.

\begin{figure}
\centering
\includegraphics{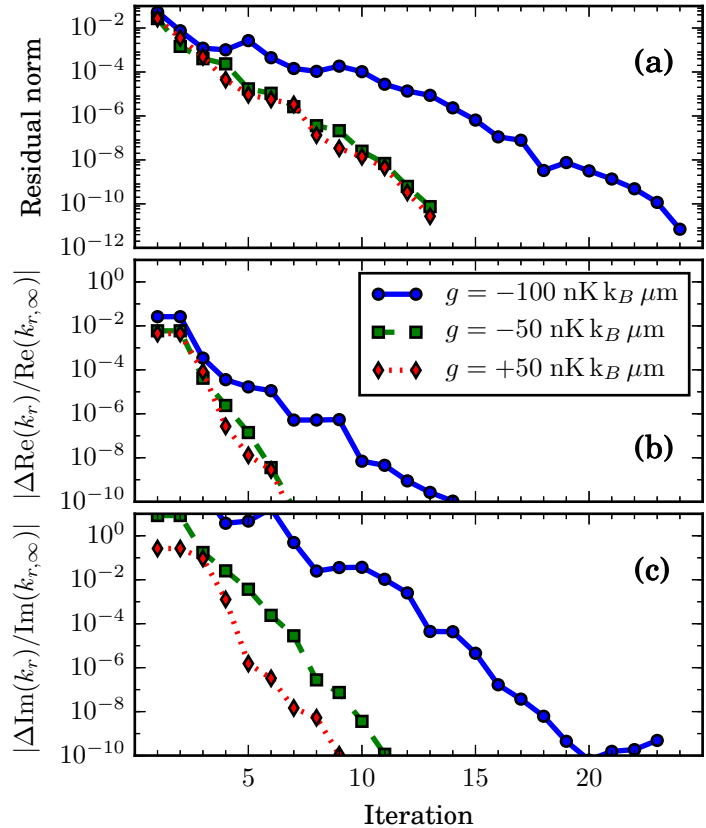}
\caption{\label{fig:davidson}Convergence of the resonance state for
  three different interaction strengths using the Davidson method. The
  model space dimension is 57600, corrsponding to
  $N_\mathrm{pts}=240$. Panel (a): Residual norm of the
  eigenstate. Panel (b): Relative convergence (distance from converged
  value) of the real part of the eigenvalue. Panel (c): Same as (b)
  but for the imaginary part of the eigenvalue. The converged
  eigenvalue is denoted $k_{r,\infty}$.}
\end{figure}

However, there exists a simpler method to distinguish these states
from the continuum of many-body scattering solutions. The resonance state is usually the
state with the largest overlap (in modulus) with $|\Phi^{(0)}\rangle$,
referred to as the pole approximation~\cite{Michel:2002hk}.
With the aim of targeting this state we employ the Davidson algorithm
for diagonalization~\cite{Davidson:1975db,Michel:606452}. The Davidson
method is very efficient at finding eigenvalues for diagonally
dominant matrices. The basic idea of this algorithm is that a search
space can be constructed by targeting a certain eigenpair. An
approximation $\left( | {\phi}_k \rangle,\hat{E}_k \right) $ for the
desired eigenpair $\left( | {\phi} \rangle ,E\right)$ is constructed
in a subspace of dimension $k$ that is much smaller than the full
dimension.  The search space is extended iteratively, as in many other
methods, but the Davidson algorithm does not rely on a Krylov
subspace.  At each iteration, we select the Ritz pair $\left( |
  {\hat{\phi}_k} \rangle ,\hat{E}_k\right)$ that has the largest
overlap with the pole approximation, $|\Phi^{(0)}\rangle$. The search
space is then extended in the direction of the residual vector $|
{r_k} \rangle = H | {\hat{\phi}_k} \rangle - \hat{E}_k |
{\hat{\phi}_k} \rangle$.  Convergence is achieved when the norm of the
residual vector approaches zero.
The main computational cost of the method is the matrix-vector
multiplication that is required at each iteration. For the problems
that we consider here, convergence is usually reached within 10-20
iterations as demonstrated in Fig.~\ref{fig:davidson}.

\paragraph{Many-body tunneling rate} 
%
Concerning the tunneling rate we want to stress that there is {\it a
  priori} no simple relation between the decay width and the half life
for a many-body system, contrary to the case of a \sp{} Gamow
state. Assuming exponential decay we would estimate the tunneling rate
$\gamma_\Gamma = \Gamma_r / \hbar = -2\mathrm{Im}(E_\mathrm{res}) / \hbar$.
However, having access to the resonant wave function,
$\Phi_\mathrm{res}(x_1,x_2) \equiv \Phi_\mathrm{res}(\mathbf{x})$, we
can alternatively compute the decay rate using an integral
formalism~\cite{Grigorenko:2007ja}. The rate of particle emissions can
be obtained by integrating the outward flux of particles at large
distance ${x_\mathrm{out}}$ from the center of the trap, and
normalizing by the number of particles on the inside
\begin{equation}
\begin{split}
\gamma_{\mathrm{flux}} = 
\frac{\hbar}{i m N({{x_\mathrm{out}}})}
\sum_i \int_0^{{x_\mathrm{out}}} \prod_{j \neq i} dx_j
\left[ \Phi_\mathrm{res}^{*}(\mathbf{x})
  \frac{d}{dx_i}\Phi_\mathrm{res}(\mathbf{x}) \right. \\ 
 \left. -\left( \frac{d}{dx_i}\Phi_\mathrm{res}^{*}(\mathbf{x}) \right)
  {\Phi_\mathrm{res}(\mathbf{x})} \right]_{x_i={{x_\mathrm{out}}}},  \label{eq:tunnel} 
\end{split}
\end{equation}
with $N({x_\mathrm{out}}) = \int_{0}^{x_\mathrm{out}} \prod_j dx_j
|\Phi_\mathrm{res}(\mathbf{x})|^2$.
%
\section{Results}
\label{sec:results}
%
In this \paper{} we restrict ourselves to the simplest instance of the
described tunable open quantum system, the case of two interacting
fermions in different spin states in an open 1D potential
trap. However, we want to stress that the formalism can be applied to
higher-dimensional traps and to systems with more particles.
For comparison with experimental
results we will use molecular units, in which energy is given in
$\unit{nK \, k_B}$, time in $\unit{\mu s}$, and distances in
$\unit{\mu m}$. In these units we have $\hbar =  \unit[7638.2]{nK \, k_B \,
\mu s}$, the Bohr magneton $\mu_B = \unit[6.7171 \cdot 10^{8}]{nK \, k_B \,
T^{-1}}$  and $\hbar^2/m = \unit[80.645]{nK \, k_B \, \mu m^2}$, where $m$
is the mass of a ${}^6$Li atom. 

In Fig.~\ref{fig:OneDContour}(a) we show for illustrative purpose the
trap potential with $p V_{0}=\unit[2.123 \cdot 10^{3}]{nK \, k_B}$,
$x_R=\unit[9.975]{\mu m}$, $B'=\unit[18.90 \cdot 10^{-8}]{T \, \mu m^{-1}}$,
$c_{B ,\sigma}=1 $, which closely resemble the parameters extracted from
experimental data (see also discussion below).
In order to handle the linear term $B' x$ we truncate the potential at
$x_\mathrm{cut}$, sufficiently far away from the relevant trap
region. In practice, this is achieved by applying a positive energy
shift $E_\mathrm{shift}$ so that $V(x_\mathrm{cut}) + E_\mathrm{shift}
= 0$. The energy shift is subtracted at the end, and we have verified
that the fluctuations in the \sp{} energy (tunneling rate) with the choice of
$E_\mathrm{shift}$ was less than $\unit[0.04]{\%}$ ($\unit[2]{\%}$).

The \sp{} Schr\"odinger equation is solved using the method described
above. The discrete set of complex-momentum states $\{
\ket{k_i} \}$ that span the contour $L^{+}$ is shown as blue dots in
Fig.~\ref{fig:OneDContour}(b). The 
energy shift that was used is $E_\mathrm{shift} =  \unit[500]{nK \,
  k_B}$.
The resulting set of eigenstates (green circles) lies very close to
the contour with the exception of one isolated state. The former
states correspond to non-resonant scattering solutions, while the
latter is a resonance.  Together, these eigenstates form the complete
set of \sp{} basis states, $\{ \ket{u_i} \}$, that will be used in the
many-body calculation.

The number of points on the contour is increased until convergence of
the \sp{} resonance energy is achieved.
Note that the resonance pole will
always remain fixed while the set of scattering states will depend on
the choice of the contour $L^{+}$. For illustrative purposes, the
contour shown in Fig.~\ref{fig:OneDContour} consists of only
$N_\mathrm{pts}=100$ basis states while full calculations were
performed with $N_\mathrm{pts} = \text{240--320}$.
For this set of potential parameters we find $e_\mathrm{res} = (301.415 -0.08548 i)
\; \unit{nK \, k_B}$, which translates into a tunneling rate $\gamma_1 =
22.38 \; s^{-1}$.

\begin{figure}
\centering
\includegraphics{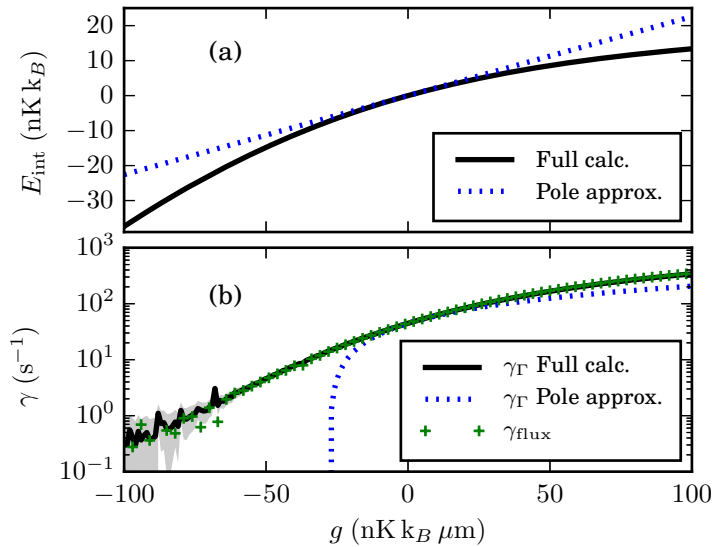}
\caption{\label{fig:ReERateStrength}Two-fermion resonance state as a
  function of the interaction strength $g$ for $c_{B,\sigma} = 1$. Panel (a):
  Interaction energy~\eqref{eq:Eint} compared with the corresponding
  energy obtained using the pole approximation. Panel
  (b): Tunneling rates obtained from the imaginary part of the
  resonance energy (from the full calculation and the pole
  approximation, respectively) compared with the rate obtained from
  the flux calculation~\eqref{eq:tunnel}. }
\end{figure}

We now consider the solution of the interacting two-fermion system,
projected on the full Berggren basis. We define the interaction energy as
\begin{equation}
E_\mathrm{int} \equiv \mathrm{Re}(E_\mathrm{res}) - 2 e_r ,
\label{eq:Eint}
\end{equation}
where $\mathrm{Re}(E_\mathrm{res}) = E_r$ is the real part of the resonance energy.
Results for the two-particle resonance state as a function of the
interaction strength $g$ are shown in Fig.~\ref{fig:ReERateStrength}.
For $g=0$, the two fermions tunnel out independently and the tunneling
rate is equal to $\gamma = 2 \gamma_1 = 44.76~s^{-1}$. However, as the
interaction becomes more attractive, the real part of the resonance
energy decreases, and the effective barrier seen by the
two particles increases. As a consequence the tunneling rate
decreases as seen in Fig.~\ref{fig:ReERateStrength}(b).

Along with the full calculations, we show in
Fig.~\ref{fig:ReERateStrength} also results obtained in the pole
approximation, which corresponds to the single configuration where the
two distinguishable fermions occupy the \sp{} resonant state. This
comparison clearly demonstrates the importance of continuum
correlations. The resonance energy and width are both decreased due to
configuration mixing between the \sp{} resonance pole and non-resonant
scattering states. In particular, the energy width, which translates
into a decay rate, is very sensitive to these correlations. These
results highlight the importance of properly taking the openness of
the system into account.

The agreement between the tunneling rate computed from the decay width
of the resonance and from the flux formula~\eqref{eq:tunnel}
demonstrates the quality of our numerical approach. It also shows that
the tunneling is well approximated by an exponential decay law for
this system.
%
\section{Numerical convergence}
\label{sec:numerics}
\begin{figure}[t]
    \centering
    \includegraphics[width=0.8\columnwidth]{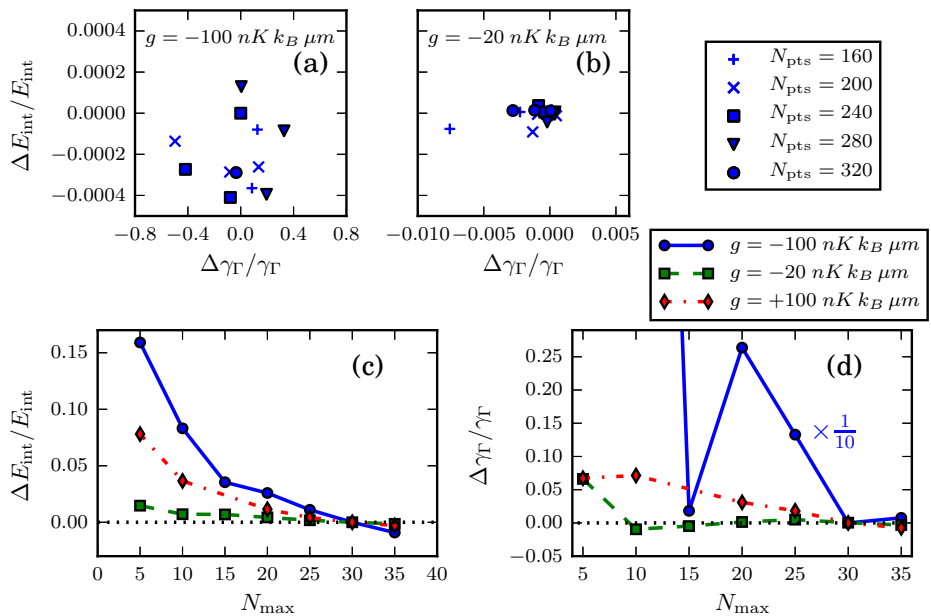}
    \caption{\label{fig:convergence}%
      Convergence study for the two-body problem with different
      interaction strengths. The upper row shows relative changes of
      the resonance pole
      position with different contours and different number of
      dicretization points, for strong attraction (panel a) and
      intermediate attraction (panel b). The two lower rows show the
      relative change of the real part (panel c) and imaginary part (panel d)
      of the resonance energy with increasing HO truncation in the
      calculation of two-body matrix elements~\eqref{eq:tbme}.
    }
\end{figure}

%
The stability of the results for the resonance energy, with respect to
different model-space parameters, may be investigated in order to
assess the numerical precision of the method. We have performed a
series of such convergence studies for systems with different
interaction strengths ($g = \unit[+100, -20, -100]{nK \, k_B \, \mu
m}$).  Panels (a) and (b) of Fig.~\ref{fig:convergence} show the
relative change of the resonance energy position for different
contours $L^+$ and different number of discretization points. Panels
(c) and (d) of Fig.~\ref{fig:convergence} demonstrate the convergence
with increasing HO truncation in the computation of interaction matrix
elements~\eqref{eq:tbme}.
Based on these results we quantify the numerical precision for a
specific interaction strength by adding (in quadrature) the amplitudes
of variations when the model-space parameters were altered one by
one. We found that the precision of the real part of the interaction
energy was on the order of $\lesssim \unit[2]{\%}$ for the entire
range of interaction strengths.
However, the precision of the computed imaginary energy was found to have a
lower bound since variations of the computed decay rate was never
smaller than \unit[0.5]{$s^{-1}$}. 
This becomes obvious when the interaction is strongly attractive and
the absolute value of the decay rate is ($\sim$\unit[1]{$s^{-1}$}),
see Fig.~\ref{fig:convergence}(a).

The larger relative variations in the tunneling rates compared to the
interaction energies, as demonstrated in this section, are most likely
due to the fact that the imaginary part of the energy is several
orders of magnitude smaller than the real part. This means that in
order to get precise results for the tunneling rates, an even more
precise result for the modulus of the energy is needed.
The estimated uncertainties from these numerical studies are shown as
shaded bands in both panels of Fig.~\ref{fig:ReERateStrength}, but is
only visible in the tunneling rate for the most attractive
interactions ($g \lesssim \unit[-60]{nK \, k_B \, \mu m} $).
%
\section{Concluding remarks}
\label{sec:conclusion}
The tunneling of few fermions from low-dimensional traps were measured
by \citet{2013PhRvL.111q5302Z}. 
The analysis of data from this experiment is quite complicated and
involves the use of the WKB approximation to extract the trap
potential parameters. More precisely, $p V_0$ and $B'$ in
Eq.~\eqref{eq:Vexp} were adjusted such that the \sp{} tunneling rates
obtained in the WKB approximation matched the experimental results.
We studied this experiment in a recent
publication~\cite{2014arXiv1412.7175L} using the method outlined in
this \paper{}.
Using the set of parameters given in Ref.~\cite{2013PhRvL.111q5302Z}
as input to our exact diagonalization approach we found a good agreement
for the \sp{} energies (with a difference of at most a few percent),
while \sp{} tunneling rates were almost two times larger than the
ones published in Ref.~\cite{2013PhRvL.111q5302Z}.
The main conclusion from these findings is that that the WKB
method should not be expected to produce reliable estimates for the
tunneling rate and that the analysis of experimental results for open
quantum systems is highly sensitive to the determination of trap
parameters.
%
\begin{acknowledgements}
  The first author wishes to acknowledge the organizers and participiants
  of the 7th International Workshop on the \emph{``Dynamics of
    Critically Stable Quantum Few-Body Systems''} in Santos, Brazil.  The
  cross-disciplinary theme of the meeting provided a very stimulating
  environment and triggered many interesting discussions.  The
  research leading to these results has received funding from the
  European Research Council under the European Community's Seventh
  Framework Programme (FP7/2007-2013) / ERC grant agreement
  no.~240603, and the Swedish Foundation for International Cooperation
  in Research and Higher Education (STINT, IG2012-5158). The
  computations were performed on resources provided by the Swedish
  National Infrastructure for Computing (SNIC) at High-Performance
  Computing Center North (HPC2N) and at Chalmers Centre for
  Computational Science and Engineering (C3SE).  We thank the European
  Centre for Theoretical Studies in Nuclear physics and Related Areas
  (ECT*) in Trento, and the Institute for Nuclear Theory at the
  University of Washington, for their hospitality and partial support
  during the completion of this work. We are much indebted to
  D. Blume, M. Zhukov, N. Zinner, and G. Z\"urn for stimulating
  discussions.
\end{acknowledgements}

\bibliographystyle{spbasic}
\bibliography{1dtunneling,1dtunneling-temp,1dtunneling-extra}

\begin{thebibliography}{20}
\providecommand{\natexlab}[1]{#1}
\providecommand{\url}[1]{{#1}}
\providecommand{\urlprefix}{URL }
\expandafter\ifx\csname urlstyle\endcsname\relax
  \providecommand{\doi}[1]{DOI~\discretionary{}{}{}#1}\else
  \providecommand{\doi}{DOI~\discretionary{}{}{}\begingroup
  \urlstyle{rm}\Url}\fi
\providecommand{\eprint}[2][]{\url{#2}}

\bibitem[{Berggren(1968)}]{Berggren:1968zz}
Berggren T (1968) {On the use of resonant states in eigenfunction expansions of
  scattering and reaction amplitudes}. Nucl Phys A109:265--287

\bibitem[{Chin et~al(2010)Chin, Grimm, Julienne, and Tiesinga}]{Chin:2010kl}
Chin C, Grimm R, Julienne P, Tiesinga E (2010) {Feshbach resonances in
  ultracold gases}. Rev Mod Phys 82(2):1225--1286

\bibitem[{Davidson(1975)}]{Davidson:1975db}
Davidson ER (1975) {Iterative Calculation of a Few of Lowest Eigenvalues and
  Corresponding Eigenvectors of Large Real-Symmetric Matrices}. J Comput Phys
  17(1):87--94

\bibitem[{Fossez et~al(2013)Fossez, Michel, Nazarewicz, and
  P{\l}oszajczak}]{Fossez:2013fj}
Fossez K, Michel N, Nazarewicz W, P{\l}oszajczak M (2013) {Bound states of
  dipolar molecules studied with the Berggren expansion method}. Phys Rev A
  87(4):042,515

\bibitem[{Gamow(1928)}]{1928ZPhy...51..204G}
Gamow G (1928) {Zur Quantentheorie des Atomkernes}. Zeitschrift f{\"u}r Physik
  51(3):204--212

\bibitem[{Grigorenko and Zhukov(2007)}]{Grigorenko:2007ja}
Grigorenko LV, Zhukov MV (2007) {Two-proton radioactivity and three-body decay.
  III. Integral formulas for decay widths in a simplified semianalytical
  approach}. Phys Rev C 76(1):014,008

\bibitem[{Hagen et~al(2006)Hagen, Hjorth-Jensen, and Michel}]{Hagen:2006fy}
Hagen G, Hjorth-Jensen M, Michel N (2006) {Gamow shell model and realistic
  nucleon-nucleon interactions}. Phys Rev C 73(6):064,307

\bibitem[{{Lundmark} et~al(2014){Lundmark}, {Forss{\'e}n}, and
  {Rotureau}}]{2014arXiv1412.7175L}
{Lundmark} R, {Forss{\'e}n} C, {Rotureau} J (2014) {Tunneling Theory for
  Tunable Open Quantum Systems of Ultracold Atoms in One-Dimensional Traps}.
  ArXiv e-prints \eprint{1412.7175}

\bibitem[{de~la Madrid(2005)}]{delaMadrid:2005gt}
de~la Madrid R (2005) {The role of the rigged Hilbert space in quantum
  mechanics}. Eur J Phys 26(2):287--312

\bibitem[{Michel et~al(2002)Michel, Nazarewicz, P{\l}oszajczak, and
  Bennaceur}]{Michel:2002hk}
Michel N, Nazarewicz W, P{\l}oszajczak M, Bennaceur K (2002) {Gamow Shell Model
  Description of Neutron-Rich Nuclei}. Phys Rev Lett 89(4):042,502

\bibitem[{Michel et~al(2003)Michel, Nazarewicz, Ploszajczak, and
  Okolowicz}]{Michel:606452}
Michel N, Nazarewicz W, Ploszajczak M, Okolowicz J (2003) {Gamow shell model
  description of weakly bound nuclei and unbound nuclear states}. Phys Rev C
  67(5)

\bibitem[{Michel et~al(2009)Michel, Nazarewicz, P{\l}oszajczak, and
  Vertse}]{Michel:2009jg}
Michel N, Nazarewicz W, P{\l}oszajczak M, Vertse T (2009) {Shell model in the
  complex energy plane}. J Phys G 36(1):013,101

\bibitem[{Newton(1960)}]{1960JMP.....1..319N}
Newton RG (1960) {Analytic Properties of Radial Wave Functions}. J Math Phys
  1:319--347

\bibitem[{Olshanii(1998)}]{Olshanii:1998jr}
Olshanii M (1998) {Atomic Scattering in the Presence of an External Confinement
  and a Gas of Impenetrable Bosons}. Phys Rev Lett 81(5):938--941

\bibitem[{Papadimitriou et~al(2011)Papadimitriou, Kruppa, Michel, Nazarewicz,
  Ploszajczak, and Rotureau}]{Papadimitriou:2011ik}
Papadimitriou G, Kruppa AT, Michel N, Nazarewicz W, Ploszajczak M, Rotureau J
  (2011) {Charge radii and neutron correlations in helium halo nuclei}. Phys
  Rev C 84(5):051,304

\bibitem[{Rotureau and van Kolck(2013)}]{2013FBS....54..725R}
Rotureau J, van Kolck U (2013) {Effective Field Theory and the Gamow Shell
  Model. The $^6$He Halo Nucleus}. Few-Body Syst 54(5):725--735

\bibitem[{Rotureau et~al(2006)Rotureau, Michel, Nazarewicz, P{\l}oszajczak, and
  Dukelsky}]{Rotureau:2006fo}
Rotureau J, Michel N, Nazarewicz W, P{\l}oszajczak M, Dukelsky J (2006)
  {Density Matrix Renormalization Group Approach for Many-Body Open Quantum
  Systems}. Phys Rev Lett 97(11):110,603

\bibitem[{Serwane et~al(2011)Serwane, Z{\"u}rn, Lompe, Ottenstein, Wenz, and
  Jochim}]{Serwane:2011hp}
Serwane F, Z{\"u}rn G, Lompe T, Ottenstein TB, Wenz AN, Jochim S (2011)
  {Deterministic Preparation of a Tunable Few-Fermion System}. Science
  332(6027):336--338

\bibitem[{Z{\"u}rn et~al(2013{\natexlab{a}})Z{\"u}rn, Lompe, Wenz, Jochim,
  Julienne, and Hutson}]{Zuern:2013cr}
Z{\"u}rn G, Lompe T, Wenz AN, Jochim S, Julienne PS, Hutson JM
  (2013{\natexlab{a}}) {Precise Characterization of Li-6 Feshbach Resonances
  Using Trap-Sideband-Resolved RF Spectroscopy of Weakly Bound Molecules}. Phys
  Rev Lett 110(13):135,301

\bibitem[{Z{\"u}rn et~al(2013{\natexlab{b}})Z{\"u}rn, Wenz, Murmann,
  Bergschneider, Lompe, and Jochim}]{2013PhRvL.111q5302Z}
Z{\"u}rn G, Wenz AN, Murmann S, Bergschneider A, Lompe T, Jochim S
  (2013{\natexlab{b}}) {Pairing in Few-Fermion Systems with Attractive
  Interactions}. Phys Rev Lett 111(17):175,302

\end{thebibliography}

\end{document}